\documentclass[10pt, conference, compsocconf]{IEEEtran}

\usepackage{amsmath}
\usepackage{microtype}
\usepackage{algorithm,algpseudocode}
\usepackage[style=ieee,url=false,doi=false,isbn=false,sorting=none]{biblatex}
\usepackage{filecontents}
\usepackage{pgfplots,pgfplotstable,booktabs}

\title{Fast Algorithm for N-2 Contingency Problem}
\author{\IEEEauthorblockN{K.~S. Turitsyn, P.~A. Kaplunovich}
\IEEEauthorblockA{Department of Mechanical Engineering\\
Massachusetts Institute of Technology\\
Cambridge, 02139, USA\\
Email: turitsyn@mit.edu, pekap@mit.edu}

}

\begin{document}
\maketitle
%\vskip0.375in
%\begin{center}
%{\fontsize{14}{17} \textbf{Fast Algorithm for N-2 Contingency Problem}}
%\end{center}
%\vskip2\baselineskip
%\begin{multicols}{2}
	\begin{center}
		{\large\bfseries Abstract}
	\end{center}
	\textit{
	We present a novel selection algorithm for $N-2$ contingency analysis problem. The algorithm is based on the iterative bounding of line outage distribution factors and successive pruning of the set of contingency pair candidates. The selection procedure is non-heuristic, and is certified to identify all events that lead to thermal constraints violations in DC approximation. The complexity of the algorithm is $O(N^2)$ comparable to the complexity of $N-1$ contingency problem. We validate and test the algorithm on the Polish grid network with around $3000$ lines. For this test case two iterations of the pruning procedure reduce the total number of candidate pairs by a factor of almost $1000$ from $5$ millions line pairs to only $6128$.
	}
	\section{Introduction}
 Maintaining reliable operation operation of the power system is of paramount importance for the power grid operators and society as a whole. This task will likely become even more challenging due to combination of multiple factors, that include shift toward intermittent renewable generation, electric transportation systems, deregulation of energy markets. The standards developed by North American Electric Reliability Corporation \cite{NERC2009} necessitate the operators to ensure the system performance in the events of multiple outage contingencies. However, the problem of contingency identification remains computationally challenging due to combinatorial explosion of the total number of possible initiating events. This number grows approximately as $N^k$ where $N$ is the number of components (typically branches of the network) and $k$ is the number of outaged elements.

	Large number of algorithms have been developed to address the problem of computational complexity. The classical approaches towards contingency identification are based on ranking and selection approaches \cite{Ejebe1979, Mikolinnas1981, Irisarri1981, Enns1982, Stott1985, Ejebe1988}. Within the ranking framework the candidate outage configurations are ranked according to heuristic performance index based on the line flow, capacity as well as the total number of lines in the network. Multiple variations of the method exist differing in the functional form of the performance index. The selection approach \cite{Mikolinnas1981, Ejebe1988} is based on the analysis of power flow solutions and provide more accurate ranking at the expense of additional computational burden. A number of modifications to both methods have been proposed in the recent years that have significantly improved the efficiency of the ranking procedure. These include the approaches based on the network topology analysis \cite{Chen2005, Guler2007, Dosano2009}, nonlinear optimization heuristics \cite{Mori2001, Donde2008, Eppstein2011} and others. Our work is most closely related to the approaches based on the Line Outage Distribution Factors that have been recently explored in \cite{Davis2009,Davis2011}.
	
	In this paper we develop a new approach towards contingency selection problem that is based on iterative pruning of the contingency candidate set. Starting with a set of all possible $2$ line outage pairs we exclude the pairs that are guaranteed to be ``safe'' from the contingency perspective. The corresponding guarantees can be shown using the analytic bounds for the line overload expression based on the Line Outage Distribution Factors computed within the stage of $N-1$ contingency analysis. For realistic cases with small number of contingencies this pruning procedure allows one to filter out most of the line combinations leaving only few potentially dangerous ones. If the number of the final candidates is  $O(N)$ or lee they can be analyzed directly with negligible computational overhead. Unlike most of the other approaches, our algorithm is not based on any uncontrolable heuristics. It is guaranteed to capture all the dangerous events without missing any pairs leading to violations. In this manuscript we describe the algorithm for $N-2$ contingency analysis, its extensions to more general $N-k$ problem will be reported elsewhere. The overall complexity of the algorithm depends on the efficiency of the power flow solution procedure and on the total amount of contingencies violating thermal constraints. In the relatively unstressed situations when the total number of contingencies is small the complexity can be estimates as $O(RN)$ where $R$ is the number of operations required to solve the linear power flow equations. The overall complexity is therefore comparable to the $N-1$ contingency problem that is routinely solved by system operators.
	
	The structure of this paper is the following. In section \ref{setting} we formally define the problem and derive the key relations necessary for the constuction of the algorithm. In section \ref{algorithm} we describe the actual algorithm and discuss the issues of complexity, implementation and possible optimizations. Next, in section \ref{results} we present the results of algorithm validation and various tests on the $3000$ bus Polish grid model. Finally, the overview of the approach as well as possible extensions and research directions are presented in section \ref{conclusions}.

	\section{Problem setting} \label{setting}
	In this work we limit ourselves to DC approximation which is also used in most of the other $N-k$ contingency studies. Although it's accuracy can be limited in some situations it is  a reasonable model for an already challenging $N-k$ contingency problem. Within this approximation the state of the power system is described by the vector of voltage phases $\theta_k$ defined on every of the $M$ buses in the system. The power flows are described by the linear dc power flow equations:
	\begin{align}\label{main}
		\mathbf{B}\theta = p
	\end{align}
	where $\hat{B}$ is the  $M\times M$ nodal DC susceptance matrix and $p$ is the vector of active power injections. The nodal DC susceptance matrix can be represented as $\mathbf{B} = \mathbf{M}\mathbf{Y}\mathbf{M}^T$, where $\mathbf{Y}$ is the diagonal $N\times N$ matrix of branch susceptances, and $\mathbf{M}$ is the $M\times N$ connection matrix with $1$s indicating the beginning bus of every branch, and $-1$ its end. The vector of power flows can be represented as $f = \mathbf{Y}\mathbf{M}^T\theta =  \mathbf{Y}\mathbf{M}^T\mathbf{B}^{-1} p$.
	
	Linear DC power flow admit a very simple and elegant analysis of the single and multiple line contingencies. There is conservation of total power flowing through the system, so whenever one or multiple line outage, the power that was flowing through them is distributed between the other lines in the system. Linear structure of the equations allows one to describe this distribution via linear mapping. The effect of the outage can be described by the matrix of so called Line Outage Distribution Factors (LODF) denoted as $L_{yx}$ that relates the change of flow in a monitored line $y$ that follows after the tripping of line $x$ with original flow $f_x$. Formally one can write:
	\begin{align}\label{LODF}
		L_{yx} = \frac{f_y'-f_y}{f_x}
	\end{align}
	relates the change of the flow through line $y$ from $f_y$ to $f_y'$ with the flow $f_x$ through line $x$ before the outage. %The conservation of the current implies that $L_{xx} = -1$ and $\sum_x L_{xy} = 0$.
	The LODFs are extensively used for the $N-1$ contingency analysis. They can be computed in $O(N K)$ operations, which is an acceptable overhead on top of the amount of calculations required to solve power flow equations. In the following discussion we assume that the matrix $L_{xy}$ has been precomputed. As we will show, it is possible to express the overload effect of the double outage in terms of the expression for single outage LODF. This relation forms the basis of our algorithm that efficiently utilizes the information available from $N-1$ contingency analysis to identify a tight set of double outage contingency candidates.
	
	In order to find the relation between single and two line contingency LODFs we use the well-known expression for the LODF in general $k$-line contingency situation (see e.g. \cite{Stott1985}):
	\begin{align}\label{Lmat}
		\mathbf{L} = \mathbf{Y}\mathbf{M}^T \mathbf{B}^{-1}\mathbf{\tilde M}(\mathbf{1} - \mathbf{\tilde Y}\mathbf{\tilde M}^T \mathbf{B}^{-1}\mathbf{\tilde M})^{-1},
	\end{align}
	where $\mathbf{\tilde M}$ is the $M\times k$ submatrix of $\mathbf{M}$ corresponding to the outaged lines and similarly $\mathbf{\tilde Y}$ is the $k\times k$ outaged line submatrix of $\mathbf{Y}$. This expression is applicable both to single ($n=1$) and double $n=2$ line outage events. Direct comparison of these expressions allows us to relate the two. LODF matrices. After straightforward but bulky calculations we arrive at the following expression for the effect of double outage:
	\begin{align} \label{fz}
		f_z' - f_z = \frac{L_{zx}(f_x + L_{xy} f_y)}{1-L_{yx}L_{xy}} + \frac{L_{zy}(f_y + L_{yx} f_x)}{1-L_{yx}L_{xy}}.
	\end{align}
	In this relation we denote the outage lines by $x,y$ and consider the change of the flow on some arbitrary line $z$. The expressions $L_{xy}$ correspond to the single line outage as defined in \eqref{LODF}. Similar expression, although written in a different form has been recently derived in \cite{Davis2011}. For some combinations of intially tripped lines $x,y$ the denominator $1-L_{xy}L_{yx}$ can be zero. It was shown in \cite{IslandingThroughZero} that such situations correspond to the islanding of the grid. After the grid is islanded the rank of the matrix $\mathbf{B}$ in \eqref{main} is increased and it may not have a solution. This corresponds to the situation when individual islands do not have balanced generation and consumption. The restoration of the balance depends on the system operator policies and is not considered in this work. In our algorithm we substitute the corresponding elements of the matrix $A_{xy}$ with zeros which automatically removes them from consideration. There are only few of such cases in the model of Polish Grid studied in this work. All of them correspond to islanding of single buses. The important property of \eqref{fz} that is extensively exploited in our algorithm is the factorization of individual terms in \eqref{fz}. After introduction of $A_{xy} = (1 + L_{xy}f_y/f_x)/(1-L_{yx}L_{xy})$ the expression \eqref{fz} can be rewritten as
	\begin{align}
	 f_z' - f_z = A_{xy} L_{zx} f_x  + A_{yx} L_{zy} f_y
	\end{align}
	The contingency occurs whenever the absolute value of the flow at line $z$ exceeds a critical value, i.e. $f_z' > f_z^{\mathrm{crit}}$ or $f_z' < - f_z^{\mathrm{crit}}$. Both of these conditions can be rewritten in the form
	\begin{align}\label{AB}
		A_{xy} B_{xc} + A_{yx} B_{yc} > 1
	\end{align}
	where the $c$ indicates one of the flow constraints, and there are two values of $c$ associated with each line $z$ with the matrix values given by $B_{xc} = f_x L_{zx}/(f_z^{\mathrm{crit}} \pm f_z)$, where the $+,-$ signs correspond to the conditions $f_z' < - f_z^{\mathrm{crit}}$ and $f_z' > f_z^{\mathrm{crit}}$ respectively. The form \eqref{AB} is rather general, and can be used for other types of linear constraints, such as voltage bus ones. Although these constraints are not discussed in this work, in the following we will assume that the sets of constraints and lines are separate and the elements of the matrix $B_{xc}$ are not necessarily associated with individual line overloads. We denote the set of possible constraints $c$ by $\mathcal{C}$ and the set of all lines by $\mathcal{E}$. In these notations the problem is reduced to selection of all tuples $(x,y)$ with $x,y \in \mathcal{E}$ such that $1-L_{xy}L_{yx} \neq 0$ for which there exists at least one constraint $c \in \mathcal{C}$ that satisfies the condition of line overload:
	\begin{align}\label{constraint}
		\Gamma_{xyc} + \Gamma_{yxc} > 1
	\end{align}
where $\Gamma_{xyc} = A_{xy}B_{xc}$. Brute force search of all such tuples requires in the worst case scenario requires at least $O(N^2 K)$ operations where $N = |\mathcal{E}|$ is the number of branches and $K = |\mathcal{C}|$ is the total number of constraints. If the only constraints are associated with line overloads $K=2N$. The iterative pruning approach described dramatically lowers this estimate in practical situation when the total number of tuples is small. In this case the complexity of the algorithm can be estimated as $O(N K) + O(N^2)$.

	\section{Algorithm}\label{algorithm}
Our algorithm is based on the simple idea of iterative pruning of the set of initiating line candidates. The algorithm exploits the algebraic structure of the overload condition \eqref{constraint}. Although both of the terms $\Gamma_{xyc}$ and $\Gamma_{yxc}$ depend on three indices $x,y,c$, these dependence has a factorized form $\Gamma_{xyc} = A_{xy}B_{xc}$. This form admits a fast bounding procedure that results in an upper bound that depends only on two indices, for instance $\Gamma_{xyc} \leq \Gamma_{xy\star}^{\max}$. This bound can be produced by finding the minimal $B_{x\star}^{\min}$ and maximal $B_{x\star}^{\max}$ values of $B_{xc}$ for every value of $z$: $B_{x\star}^{\min} \leq B_{xc} \leq B_{x\star}^{\max}$ and can be found by direct iteration over the matrix $B_{xc}$ in only $O(|\mathcal{E}|\cdot|\mathcal{C}|)$ operations. The expression for $\Gamma_{xy\star}^{\max}$ is given by
	\begin{align}
		\Gamma_{xy\star}^{\max} = \left\{
		\begin{array}{l}
			A_{xy} B_{x\star}^{\max}, \quad A_{xy} \geq 0\\
			A_{xy} B_{x\star}^{\min}, \quad A_{xy} < 0			
		\end{array}\right.
	\end{align}
	that can be compactly written as $\Gamma_{xy\star}^{\max} = \max\{A_{xy} B_{x\star}^{\max},A_{xy} B_{x\star}^{\min}\}$. As the bound $\Gamma_{xy\star}^{\max}$ depends only on two indices, it can be used for fast pruning of the set $\mathcal{A}$ of possible $(x,y) \in \mathcal{A}$ tuple candidates. Whenever $\Gamma_{xy\star}^{\max} + \Gamma_{yx\star}^{\max} \leq 1$, the condition \eqref{constraint} can not be satisfied for any possible choice of $z$. Thus, the pruning of set $\mathcal{A}$ can be accomplished in only $O(|\mathcal{A}|)$ operations which is at most $O(N^2)$. Analogous upper bounds can be constructed for $\Gamma_{x\star c}^{\max}$ and $\Gamma_{\star yc}^{\max}$ to prune the set of pairs $x,z$ that can be part of the triple satisfying \eqref{constraint}. The detailed  algorithm is presented in three listings below. The main function \texttt{findTuples} takes the set $\mathcal{E}$ of possible initiating lines and set $\mathcal{C}$ of all the relevant constraints as an input and returns the set of possible candidate tuples $\mathcal{A}$ as the output. The pruning happens in iterative fashion as each reduction of one set produces better bounds on the matrices $A,B$ and allows extra pruning of the second set.
	\begin{algorithm}[ht]
		\caption{\texttt{findTuples($\mathcal{E},\mathcal{C}$)}}
		\begin{algorithmic}[1]
			\State $\mathcal{A} \gets \{(x,y): x,y \in \mathcal{E}\}$
			\State $\mathcal{B} \gets \{(x,c): x \in \mathcal{E}, c \in \mathcal{C}\}$ \label{initB}
			\Repeat
\label{stepBstar}				\State Calculate $B_{\star c}^{\max},B_{\star c}^{\min}$				 \Comment{Prune $\mathcal{B}$}
				\State Calculate $A_{x\star}^{\max},A_{x\star}^{\min},A_{\star y}^{\max},A_{\star y}^{\min}$
				\For{$(x,c) \in \mathcal{B}$}\label{loopB}
					\State $\Gamma_{x\star c}^{\max} \gets \max\{A_{x\star}^{\max} B_{xc},A_{x\star}^{\min}B_{xc}\}$
					\State $\Gamma_{\star yc}^{\max} \gets \max\{A_{\star y}^{\max} B_{\star c}^{\max},A_{\star y}^{\min} B_{\star c}^{\min}\}$ \label{Gsxc}
				\EndFor
				\State $\mathcal{B} \gets \{(x,c)\in \mathcal{B}: \Gamma_{x\star c}^{\max} + \Gamma_{\star xc}^{\max} > 1\}$ \label{filterB}
%				\COMMENT Prune set $\mathcal{A}$
				\State Calculate $B_{x\star}^{\min},B_{x\star}^{\max}$\Comment{Prune $\mathcal{A}$}
				\For{$(x,y) \in \mathcal{A}$}\label{loopA}
					\State $\Gamma_{xy\star}^{\max} \gets \max\{A_{xy} B_{x\star}^{\max},A_{xy} B_{x\star}^{\min}\}$
				\EndFor
				\State $\mathcal{A} \gets \{(x,y)\in \mathcal{A}: \Gamma_{xy\star}^{\max} + \Gamma_{yx\star}^{\max} > 1\}$ \label{filterA}
			\Until{ $\mathcal{A}$  stops changing} \\
			\Return $\mathcal{A}$
		\end{algorithmic}
	\end{algorithm}
In the step \ref{stepBstar} we have omitted the definition of $B_{x\star}^{\min} = \min_{(x,c)\in\mathcal{B}} B_{xc}$ and its obvious counterparts for the sake of presentation simplicity. The sets $\mathcal{A},\mathcal{B}$ can be implemented via different data structures. The simplest, although not the most efficient choice is to simply use boolean masks for the matrices $A_{xy}, B_{xc}$. In this case both the iteration over the sets $\mathcal{A},\mathcal{B}$ in lines \ref{loopB}, \ref{loopA} and the filtering operations in lines \ref{filterB}, \ref{filterA} can be implemented as a direct loop over all possible values. In this implementation the total complexity of the algorithm will be given by $O(I N K) + O(I N^2)$ where $I$ is the number of outer loop iterations. More sophisticated implementations of the sets can significantly reduce the number of inner loop iterations for small set cardinalities and thus improve the overall complexity. In general, we expect that the total number of outer loop iterations necessary for the algorithm to converge will be of order $2-4$ for the realistic situations with small number of contingencies. This observation is supported by our numerical experiments, but its formal proof is far beyond the scope of our work.

Apart from various implementation possibilities there is also an additional degree of freedom related to the definition of the matrices $A_{xy}$ and $B_{xc}$. The expression $\Gamma_{xyc} = A_{xy} B_{xc}$ is invariant under the transformation $A_{xy} \to s_x A_{xy}, B_{xc} \to s_x^{-1} B_{xc}$ for any non-zero values of $s_x$. This transformation affects the value of the bound $\Gamma_{\star xc}^{\max}$ on line \ref{Gsxc} and can be used for improving the efficiency of the pruning process. Our preliminary results indicate that it is possible to reduce the size of the final set $\mathcal{A}$ by a factor of $2$  via careful choice of $s_x$. However, this reduction comes at the expense of substantial computational overhead. Nevertheless, this optimization may become important in situations where the unoptimized pruning procedure is inefficient for some reasons.
	
It is also possible to improve the efficiency of the pruning procedure by appropriate subdivision of the constraint set $\mathcal{C}$. As the bounds $B_{x\star}^{\max}$ and others are based on the analysis of the whole set of branches, few outliers in this set can significantly affect the value of the bounds. For example, a single line $z$ with flow $f_z$ very close to the capacity $f_z^{\mathrm{crit}}$ can inflate the values of $B_{x\star}^{\max}$ for all initiating lines $x$ and thus affect the efficiency of pruning. It is possible to mitigate this problem by subdivision of the constraint set $\mathcal{C}$ and separate analysis of the outlier and all the other lines. We are currently exploring these possibilities and will report our findings in future publications.
	
\section{Results}		\label{results}
In order to validate and test the proposed algorithm we have used the Polish grid model available in MATPOWER package \cite{Zimmerman2011}. This grid consists of $3269$ lines and $2737$ buses. Our simulations have started with the base state found via solution of OPF problem. The results of $N-1$ contingency analysis for the base state indicate that there are $27$ single line outage events that cause violations of one or more constraints with overall total of $37$ $(x,c)$ event-overload pairs. In order to separate these contingencies we remove the corresponding $(x,c)$ pairs from the original $\mathcal{B}$ set after step \ref{initB} of the algorithm. In order to validate the pruning algorithm we have performed an exhaustive analysis of all possible $2$ line contingencies and found 524 pairs of lines that result in overloads. Note, that this number is significantly less than the total number of ${N(N-1)/2}\approx 5.3*10^{6}$ pairs and ${N(N-1)(N-2)/6}\approx 5.8*10^{9}$ $(x,y,c)$ triples that need to be analyzed with brute force approach.

\begin{table}[ht]
\begin{center}
\pgfplotstabletypeset[
	int detect,
	columns={iter,A,B},
	columns/iter/.style={column name=\textsc{Iteration}},
	columns/A/.style={column name=$|\mathcal{A}|$},
	columns/B/.style={column name=$|\mathcal{B}|$},
	every head row/.style={before row=\toprule,after row=\midrule},
	every last row/.style={after row=\bottomrule}
]{
iter	A	B
0	5341546	10683092
1	17928	322365
2	6128	188761
3	5816	163788
4	5750	156807
5	5750	155813
6	5750	155813
}
\end{center}
\caption{Candidate set $\mathcal{A}, \mathcal{B}$ sizes evolution with algorithm progression.}
\label{restable}
\end{table}

Our algorithm has managed to reduce the number of $(x,y)$ pair candidates from $5.3*10^{6}$ to $6128$ (that of course contain all 524 pairs that actually lead to overload) in only two steps. The subsequent outer loop iterations had marginal effect on the total number of pairs. Table \ref{restable} shows the evolution of the set $\mathcal{A}, \mathcal{B}$ sizes with each iteration. Note, that although the there are a lot of elements in $\mathcal{B}$ set, they don't affect the overall effectiveness of the approach, as the output of the algorithm consists only of the initiating pairs $(x,y)$ from the set $\mathcal{A}$. As one can see from the table, the algorithm converges after 6 iterations, but only the first two iterations lead to strong reductions in the $\mathcal{A}$ set size, whereas the consequent iterations have diminishing returns.
%A matrix
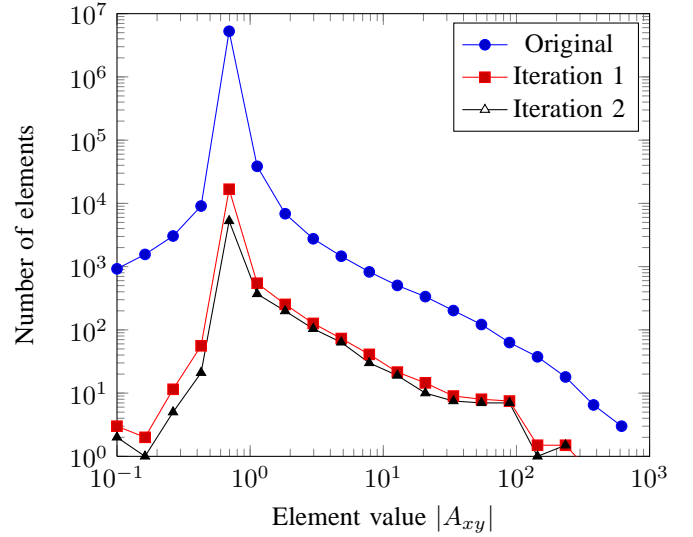
\begin{figure}[ht]
\centering
\begin{tikzpicture}
    \pgfplotstableread{histA.txt} \histAtable
	\begin{loglogaxis}[
     xlabel={Element value $|A_{xy}|$},
     ylabel={Number of elements},
     xmin=0.1, xmax=1000,
     ymin=1, ymax=1E7,
     width=\columnwidth,
%     log basis y = 100
%     ytick = {1e1,1e3,1e5,1e7},
%     yminorticks = true
     ]
		\addplot table[y=1] from \histAtable ;	
		\addplot table[y=2] from \histAtable ;	
		\addplot[mark=triangle*] table[y=3] from \histAtable ;	
		\legend{Original, Iteration 1,Iteration 2}
	\end{loglogaxis}
\end{tikzpicture}
\caption{Histogram of $A$ matrix elements distributions for the first two iterations.}
\label{Adist}
\end{figure}

In order to better understand the reason for the algorithm efficiency we have analyzed the distributions of the elements in the matrices $A_{xy}$ and $B_{xy}$. As one can see from the figure \ref{Adist} in the original system most of the elements of the matrix $A$ are close to $1$. This is because most of the lines do not affect each other after outages, so $L_{xy}, L_{yx} \ll 1$. Typically the flow from line $x$ is distributed amongst its closest neighbors, whereas most of the lines $y$ are not close in neither geographical nor electrical metrics. There are only about $10^4$ pairs in the original network with value of $A_{xy}$ larger than $1$. As expected, the pruning operations have more significant effect on the left part of the distribution, as the corresponding pairs have lower chance of producing strong overflows. The third iteration of the algorithm has a seemingly minor effect on the distribution, but this is largely an artifact of the logarithmic scale of $y$ axis, as the overall effect on the total number elements is quite significant as seen from the Table \ref{restable}.

%B matrix
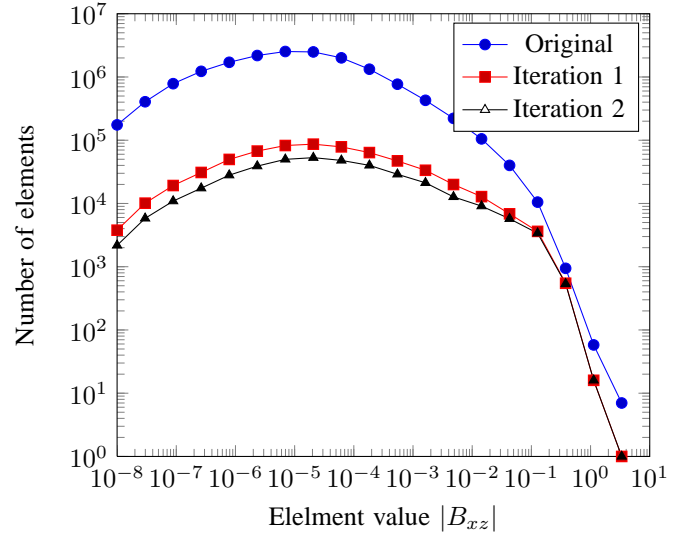
\begin{figure}[ht]
		\begin{tikzpicture}

\pgfplotstableread{histB.txt} \histAtable
	\begin{loglogaxis}[
     xlabel={Elelment value $|B_{xz}|$},
     ylabel={Number of elements},
     xmin=0.00000001, xmax=10,
     ymin=1, ymax=1E7,
     width=\columnwidth,
     ]
		\addplot table[y=1] from \histAtable ;	
		\addplot table[y=2] from \histAtable ;	
		\addplot[mark=triangle*] table[y=3] from \histAtable ;	
		\legend{Original, Iteration 1,Iteration 2}
	\end{loglogaxis}
\end{tikzpicture}
\caption{Histogram of $B$ matrix elements distributions for first two algorithm iterations.}
\label{distB}
\end{figure}

The histogram \ref{distB} of the matrix $B$ element has very different structure because the element $B_{xc}$ is proportional to the line outage distribution factor $L_{zx}$ that, as discussed previously, is very small for most of the pairs $(x,z)$. It is rather interesting that the distribution of $B_{xz}$ and $L_{xy}$ values (not shown) has an almost flat distribution in the log-scale, that points out to some self criticality in the network. We are not aware of any simple interpretations of this property. However, this property if shown to be universal for large scale power grids could be possibly linked to the power law distribution of large blackout sizes \cite{Dobson2007, Hines2009, Hines2009a} and potentially exploited for construction of fast contingency selection algorithms.

	\section{Conclusions} \label{conclusions}
	
	In conclusion, we have presented a novel algorithm for the $N-2$ contingency problem. The algorithm is based on the idea of iterative pruning of the possible candidate sets. Given the matrix of single line outage distribution factors only a small number of candidates can be identified in only $O(N^2)$ operations, much smaller than the naive exhaustive search analysis that would require $O(N^3)$ operations, therefore our algorithm decreases computational time by a factor of $O(N)$ and and its complexity is comparable with the complexity of usual $N-1$ contingency analysis Unlike many other approaches our algorithm is not heuristic, and is certified to return all the double outage with violations. The algorithm has been validated and tested on the Polish grid example where the total number of double outage with violations was shown to be $524$ via exhaustive search analysis. Our algorithm has reduced the set of all possible candidates from approximately $5000000$ to about $6000$ in just two iterations.
	
	Although the effectiveness of the approach is impressive, there are several directions one can pursue to improve it even further. First, a number of additional optimizations are possible. Apart from the optimizations and implementation discussed briefly in the end of the section \ref{algorithm}, there are a number of opportunities how this approach can be extended to more challenging settings. First, it is possible to apply the approach directly to $N-k$ problems with $k \geq 2$. This would require accurate analysis of the expression \eqref{Lmat} and derivation of relations similar to \eqref{fz}. Whenever only a small subset of possible $k$-line contingencies leads to violations, the proper bounding procedure should be able to filter out the safe candidates. Another direction is associated with extension of out approach to AC power flows. As the approach is based on bounding various contributions to the line outage distribution factors, it might be feasible to extend to nonlinear systems without having to solve them in closed form. This is certainly a much more formidable task that necessitates a rather advanced nonlinear analysis approaches.
	
	Another exciting opportunity lies in applying the proposed algorithm to the problem of analysis and mitigation of cascading failures in power grids \cite{Dobson2007, Dobson2010, Koenig2010}. The pruning approach can be used both for the development of efficient algorithms of assessing the probabilities of cascading outages, and for finding optimal decision choices for cascade prevention.

	\section*{Aknowledgements}
%\nocite{*}
This work was partialy supported by NSF award ECCS - $1128437$,  MIT/SkTech seed funding grant and RGC (Russian goverment contract) 11.519.11.6018.

	\printbibliography

\end{document}